# User's Privacy in Recommendation Systems Applying Online Social Network Data: A Survey and Taxonomy

*Erfan Aghasian[1], Saurabh Garg[2], and James Montgomery[3]*

Recommender systems have become an integral part of many social networks and extract knowledge from a user's personal and sensitive data both explicitly, with the user's knowledge, and implicitly. This trend has created major privacy concerns as users are mostly unaware of what data and how much data is being used and how securely it is used. In this context, several works have been done to address privacy concerns for usage in online social network data and by recommender systems. This paper surveys the main privacy concerns, measurements and privacy-preserving techniques used in large-scale online social networks and recommender systems. It is based on historical works on security, privacy-preserving, statistical modeling, and datasets to provide an overview of the technical difficulties and problems associated with privacy preserving in online social networks.

## 1.1 Introduction

Online social network services have become one of the most well-liked and accepted services on the Internet. These networks deliver an infrastructure so individuals can connect with one another, share information, explicit their emotions and attitudes and shape and keep a connection with different individuals on the Internet [1, 2]. These networks such as Facebook, LinkedIn, Google+, Twitter and other social networking sites all have advantages, both practical (such as sharing employment history in a LinkedIn profile) and social (like connecting with distant friends via Facebook).

In order to increase the data utility in such social networks, recommender systems can deliver personalization of a collection of items to online social network users based on their nature. Meanwhile, the personalized suggestions and recommendations in these systems are heavily dependent on users' information. This can increase the probability of information leakage of users in such networks [3]. Further, information sharing creates real threats to a user's privacy. In this case, there is a need for data protection. Fundamentally, data protection means clear sets of rules and regulations, policies and diverse measures that provided for information security and lessening the invasion into a user's privacy. This invasion can be initiated by gathering, storing and distribution of private data [4]. Hence, there is a need to understand the various types of recommender systems, privacy concerns in such systems and the ways which users' privacy can be protected.

[1]School of Technology, Environments and Design, University of Tasmania, Hobart, Australia
Email: Erfan.aghasian@utas.edu.au
[2]School of Technology, Environments and Design, University of Tasmania, Hobart, Australia
Email: Saurabh.Garg@utas.edu.au
[3]School of Technology, Environments and Design, University of Tasmania, Hobart, Australia
Email: James.Montgomery@utas.edu.au

In this survey we first introduce recommender systems and their techniques. Then, in Section 1.3, we discuss the risks and concerns for users in online social networking sites, and explain the different methods for scoring users' privacy in such networks. In Section 1.4, we describe the privacy preserving approaches. Finally, we describe privacy-preserving models in online social networks and recommender systems.

## 1.2    Recommender Systems and Techniques: Privacy of Online Social Network Data

A recommender system delivers a set of items that is pertinent to a specific user of a system [3]. This set can be vary based on the nature of each online social networking sites. Moreover, prediction in these systems is provided based on the characteristics of the user and the item itself which would be recommended to users [5]. These systems do collaborate with users and just suggest specific items which the users may be interested in [6].

Several classifications have been mentioned and considered for recommender systems [6]. The most well-known one is collaborative [7] which is applicable in different fields of research and industry. The second method is content-based. In this method, there is a need for long-term observation of users' preferences in social network [8]. The third technique is demographic. This technique is useful for the times that there is not much information about a user's preferences. Therefore, demographic information such as age or education level is used in such systems [9]. The last type is knowledge-based. Providing feedback by users is the key point for designing these type of systems. By providing more feedback to the system, the knowledge of the system will improve and better recommendations will be provided for users [10].

While these systems bring many advantages for the users in online social networking sites, the privacy risks inherent to data gathering and processing are often underrated or disregarded. In order to protect the privacy of a user's sensitive information and avoid any security concern, there is a need to apply encryption and anonymisation techniques (the common method for privacy preservation in recommender systems is encryption-based while anonymisation of data set is common for privacy-preservation of online social networks users). In this section, first, the definition of privacy is discussed to understand what is considered as sensitive or private information. Then, classification and different types of online social network are discussed to clarify the goal of each online social network sites and what can lead to privacy concern or risk for users in such sites.

### 1.2.1  Privacy: Definition

Increased utilization of information technology and communication has had a major effect on the connections between individuals. This is predominantly related to individuals who make use of transportable machines to interconnect with one another or to connect the world-wide web [11]. The word privacy has numerous subtly various definitions. These can be vary from personal privacy to information privacy from one place to another, which privacy alone is being used for multi-purpose on the Internet [12]. Table 1.1 shows some definitions for privacy and information privacy.

*Table 1.1  Definitions for privacy and information privacy*

| Definition | Authors | Year |
| --- | --- | --- |
| "*Protecting personal information from being misused by malicious entities and allowing certain authorized entities to access that personal information by making it visible to them*" | Bünnig and Cap  [13] | 2009 |
| "*An individual's claim to control the terms under which personal information identifiable to the individual is acquired, disclosed or used*" | Kang [14] | 1998 |
| "*Set of privacy policies that force the system to protect private information*" | Ni et al. [15] | 2010 |
| "*Is in disarray [and n]obody can articulate what it means*" | Solove [16] | 2006 |
| "*The ability of the individual to personally control information about one's self*" | Stone et al. [17] | 1983 |
| "*Applications that seek to protect users' location information and hide some details from others*" | Taheri et al. [18] | 2010 |
| "*Multidimensional, elastic, depending upon context, and dynamic in the sense that it varies with life experience*" | Xu et al. [19] | 2011 |

Meanwhile, the notion and idea of privacy is varied, one particular description of privacy cannot cover all characteristics of the phrase. Accordingly, based on the meaning of privacy, this study is concerned principally with information privacy of users. Regarding Kang (1998)'s definition of privacy, user's information privacy concept is intensely connected to the notation of confidentiality, which is one of the main attributes (qualities) of information Security (infosec), but not to be used in an interchangeable manner. It should be noted that confidentiality[4] is concerned with disclosure of pieces of information of an individual or its secrecy, while information privacy deals with information ownership and the consequences that information disclosure has on the individual and his/her data access permissions and controls.

### *1.2.2   Online Social Networks, Classification, and Privacy*

Popularity and interest in social networks have increased considerably over the last era. Kaplan and Haenlein [22] described social networks as applications that permit individuals to form profiles, send requests to join friends, and see other users' profiles. Many forms of information can be included in these profiles, including pictures, audio files, videos and even posts and blogs, each of which may be public or semi-public, visible to a subset of other users who use that social network site [23]. LinkedIn, Friendster, MySpace, and Facebook are among the most well-known social networks that attract many users share their information. In these sites, recommender systems can undoubtedly support to expand user participation by providing new friend recommendations or content that a user may be interested in [24].

By raising the number of users in social networks and sharing more information in these sites and also with recommender systems, concerns about users' privacy will increase. The quantity of data that social network sites collect from users is constantly growing while users' data are extremely valuable for many purposes such as research, marketing and numerous additional goals [25]. Simultaneously, an important amount of sensitive information can be obtained from users' data, which should be preserved against unapproved access and revelation [26].

As mentioned, each social network follows a different goal compared with other ones. Basically, social network sites can be categorized based on different purposes. Some social networks were founded for dating purposes, while other social networks were established for purposes such as chatting, socializing, enforcing real-life relationships and also business. Beye et al. [12] considered two main social network types and provided the purpose of use of these social networks as well.

Moreover, the functionality of each online social network site differs from the others. Based on their functionality, the social network provider may ask users different information to provide and share. Hence, information can be shared and disclosed within the different sources of social networks. Moreover, privacy concerns for users may increase as they share more and more information within different online social networks. Therefore, the privacy of users should be taken into account. Privacy includes protecting a portion of information in its scope. Three factors define this scope [12]. The first dimension is breadth, which reflects the number of groups of people. The second factor is depth, which shows the degree of allowed usage. The last factor is lifetime, which indicates the duration.

In any of these three dimensions, when a portion of data or information is moved outside the planned scope, whether maliciously or accidentally, a breach of privacy happens. Apart from the scope, users in online social networks are contending with privacy boundaries. Three boundaries have been recognized for privacy. The first boundary for individuals is disclosure. Here, users try to handle the anxiety of disclosing their information in a public or private manner. The second boundary is identity. The identity boundary is described as the ability to manage one's information with particular groups. For example, it shows users' behaviors in different situations: one at work and the other at a party. The last one is a temporal boundary. It shows how the conduct of individuals may differ over time [27].

---

[4]Confidentiality, integrity and availability are the main qualities of information security which are known as CIA-triad [20]. Confidentiality is related to unauthorized information publishing, integrity describes as modifying information in an unauthorized manner, and integrity is described as unauthorized denial of use of individual information [21].

## 1.3 Taxonomy of Privacy

Privacy can be studied from different aspects: privacy concerns, scoring models and privacy-preserving models and approaches (Figure 1.1). The first two divisions illustrate the privacy concerns in online social networks and users-related concerns, while, the third part illustrates the measurement techniques of privacy of users' data. The last two sections show the preserving approaches of privacy for users in online social networks and recommender systems.

### 1.3.1 Privacy Concerns in Social Networks

Privacy in social network sites can be seen from two different perspectives. The first perspective is local privacy or user-centric, which is known as social privacy. The second perspective is global or network-centric which is known as institutional privacy [26]. From the user-centric perspective, users decide what to share with others while they can create various levels and circles of friends, posts and information to whom they intend to share. From the global view, social network sites take advantage of users' information for different goals as stated and detailed in data usage rule and policy. Moreover, the network-centric privacy can also be seen from two distinctive approaches. Considering the first approach, the data collector is the data owner of the users' information. While these social network sites have infinite access to a user's data, the concerns for privacy are less if the data collector is trustworthy for the user. The second approach is known as surveillance privacy, where a user is suspicious about data collector. This happens when information is released to third parties by a reliable data collector [26]. Table 1.2 shows an overview of privacy concerns in social networks.

### 1.3.2 User-specific Privacy Risks and Invasion

Privacy concerns and risks for users in an online social network can be differentiated in two groups: user-related and provider-related. Users may face several issues if their information being revealed and breaches happen. This type of breach can occur by a deliberate act of hacking or the individual can disclose the data accidentally (lingering data). Considering the user-related concerns, various threats can endanger users' privacy. These threats and concerns can be disclosure of private information to strangers, inability to hide information from a certain group of friends or a friend and other users' posts about an individual [28, 29]. Another form of threat is related to the social network provider. In this case, users do not have control to preserve their privacy or shared information, specifically with other parties. Several authors investigate related threats and concerns, which can include issues with data retention, private information browsing by online social network internal employees and sell users' data to other parties [30, 31] . Although the companies apply anonymisation processes to the data before selling or sharing them with other parties, there are still other risks like re-identification that is often ignored or discounted [12]. Table 1.3 categorizes users' privacy concerns in online social networks.

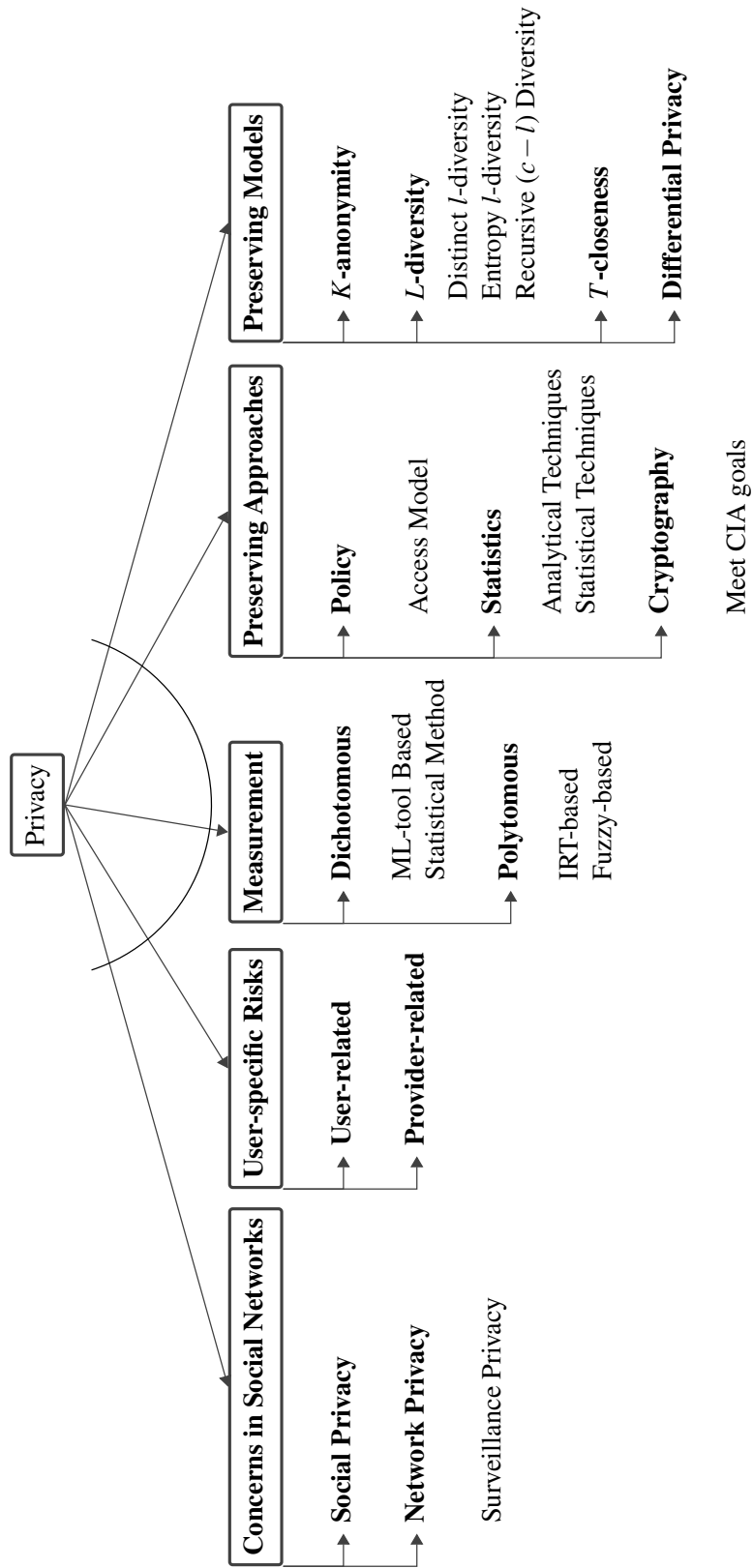

Figure 1.1: Taxonomy of Privacy in Online Social Networks

*Table 1.2   Overview of privacy concerns in social networks*

| Privacy Perspective | Related Concerns |
| --- | --- |
| Social privacy | • User awareness<br>• Complexity of privacy controls<br>• Changes in privacy controls<br>• Conflicts in privacy control |
| Network Privacy | • Use and revenue from collected data<br>• Lack of appropriate anonymisation<br>• Expand disclosing of collected data |
| Surveillance Privacy | • Unauthorized data collection<br>• Untrusted provider<br>• Non oblivion - data may be published or stored forever |

Though social-network users gain advantages from their online presence, they are often incapable of evaluating the risks to privacy which are imposed by sharing information. Even privacy-aware individuals, who care about confidentiality, seem to be eager to compromise their confidentiality to develop their digital attendance in the cybernetic environment. Users realize that control loss on private data of them results in a continuing risk. However, they are not able to analyze and evaluate the whole and long-standing threat precisely. Worse, setting privacy preferences in online services are often a complex and laborious duty that numerous individuals feel confused about and typically ignore or skip [32].

On the other hand, users share their information with others while they do not intend to. For example, it has been reported that nearly 32% of users on the Internet have experienced privacy attacks due to various types of unintentional disclosure, like private materials distributed in an unwilling manner or being tagged in a discomforting photo, reluctantly [33].

*Table 1.3   Users' information privacy concerns in online social networks*

| Privacy Perspective | Related Concerns |
|---|---|
| User-related | • Disclosure of private information for the strangers<br>• Inability to hide information from a certain group of friends or a friend<br>• Other users' posts about an individual |
| Provider-related | • Issues with data retention<br>• Private information browsing by online social networks internal employees<br>• Selling the users data to other parties |

The other point is that historical research on privacy has frequently concentrated on clarifying privacy matters related to information disclosure that may have tangible significance for affected users, such as financial information. Although publishing and disclosure of this type of information bring privacy risks in online social networks, the revelation of such data is less usual, if not uncommon [34]. In fact, there is evidence that revelation of such private embarrassing information, could disturb users' social position and their relationships [35], establish the key source of risks to users' privacy in online social networks [36].

### 1.3.3   Measuring Privacy in Online Social Networks

One of the significant tasks that should be considered in online social networks is privacy scoring and measurement. It is not inherently clear which information can result in a significant loss such as identity theft. Other risks are even harder to measure: comments and pictures of a user, which is risk-free for a number of individuals, can be detrimental to others. One case is a criticism against a religion or government. In some countries and cultures such criticism is broadly accepted, whereas, in other countries, an individual can get in severe difficulties for performing such an action [37, 38]. Another risk of using online social networks is posting vacation information when users are abroad. Therefore, intruders could decide when to rob the house based on the information they gather. There are several techniques and methods for calculating the privacy and information sharing in a public manner [38]. Different authors have proposed various techniques and methods from the algorithmic approach to statistical ones to score and measure the privacy. The main two approaches for measuring the privacy are dichotomous and polytomous. For each approach, several models has been proposed to measure the privacy of users in online social networks. The most well-known methods related to these approaches has been discussed in this study.

### A. Dichotomous Approach

In this part, first, we discuss a privacy risk formula proposed by Renner [37]. Then we explain a quantifying privacy approach proposed by Becker et al. [39]. Next, we introduce privacy awareness enhancement that was proposed by Petkos et al. [40]. It is worth noting that all current privacy-scoring methods focus on the single source of data of individuals, while users share their information in different sources of online social networks. Finally, other models and techniques for calculating and measuring privacy risk are introduced.

### 1.3.3.1 Renner Privacy Risk Formula

Renner [37] discusses a common approach for defining privacy risk by considering two privacy metrics including negative consequence information leakage and the likelihood of information leakage. This is given by:

$$Risk = Negative\ consequence \times Likelihood \tag{1.1}$$

Using such a formula in the context of an online social network will come with a problem: both consequences and likelihood are unknown. As a case in point, it is not easy to define the consequences of leaking an embarrassing picture, which could vary from embarrassment in front of friends through to job termination [41]. An even plainer metric to assess privacy is the number of individuals who can access information at a certain time. Of course, this metric can merely guarantee a definite privacy level when the number of individuals with access is adequately small. In this case, a user share the information to whom he/she recognizes and trusts. Nevertheless, this metric is regularly used in the real world. Most individuals discuss private information while on public transport as they assume that only the people in the same compartment will hear them talking, while they would never discuss the same subjects when talking into a microphone such that everyone in the public transport could hear them. In addition to the size, the arrangement of the group, likewise, needs to be taken into account. However, the simple metric of the group size still conveys a sense for how private or public some piece of information is [37].

### 1.3.3.2 Privacy Risk Score

Maximilien et al. [42] proposed a privacy score model to compute the risk of users who participate in online social networks considering the dichotomous approach. For creating the model, they considered visibility and sensitivity of users' information. By considering $\beta_k$ as the sensitivity of an attribute and $v(k,l)$ as the visibility of attribute $k$ of user $l$, the final scoring model is described as $PR(k,l)$, which is any combination of visibility multiple in sensitivity. The final equation is given by:

$$PR(k,l) = \beta_k \bigotimes v(k,l) \tag{1.2}$$

From the privacy score equation, it can be seen that there is a need to compute the visibility and sensitivity for the final measurement of users' privacy score. Hence, Maximilien et al. [42] provided formulas for sensitivity and visibility.

Maximilien et al. [42] mentioned that the sensitivity shows the difficulty of sharing an attribute to other users freely accessible. Based on his formula, Sensitivity of an attribute can be calculated given by:

$$\beta_k = \frac{(M - |R_k|)}{M} \tag{1.3}$$

Where $|R_k|$ is the number of individuals that make their attributes publicly available. As Maximilien et al. use a dichotomous approach, the final computed value for sensitivity is between [0,1], where the more sensitive attributes of a user have higher sensitivity score.

The other factor that has an impact on users' privacy is visibility, which Maximilien et al. calculate according to equation 1.4, which measure the probability that user $l$'s $k$th attribute is public.

$$P_{kl} = Prob[R(k,l) = 1] \tag{1.4}$$

### 1.3.3.3 Privacy Quotient Using a Naive Approach

One of the models for calculating a privacy score is proposed by Srivastava [43]. In order to model his datasets, by focusing on text messages, they deploy a naive privacy quotient. With $M$ individuals and $m$ attributes, they formed a response matrix ($M \times m$), which comes with a different range of attributes and users. They listed the profile items as follows: contact number, email, address, birth date, home-town, current town, job details, relationship status, interest, religious views and finally political views. Srivastava [43] dealt with a response matrix by assigning values 0 and 1 for shared information and unshared information about the profile, respectively. Their privacy quotient is capable of measuring sensitivity and visibility of the information as the main parameters of measuring the privacy. In their formula, they measure sensitivity for one user in relation to how much the other users share. For sensitivity calculation the following formula has been proposed as:

$$\beta_k = \frac{(M - |R_k|)}{M} \tag{1.5}$$

Where $|R_k| = \sum_l R(k,l)$ is the sum of all public values $k$ of a column of a profile. On the other hand, visibility has been defined by the following steps:

(1) $V(k,l) = Pr[R(k,l) = 1] \times 1 + Pr[R(k,l) = 0] \times 0;$

(2) $V(k,l) = Pr[R(k,l) = 1] \times 1 + 0;$

(3) $V(k,l) = Pr[R(k,l) = 1] \times 1$

Where $Pr[R(k,l) = 1]$ shows the probability that an attribute of a user is public, while $Pr[R(k,l) = 0]$ indicates that the attribute is private (by applying these steps, it can be identified which cell will be equal to 1 and which cell will have a zero value). The final equation for calculating the visibility is as follows:

$$V(k,l) = \frac{|R_k|}{M} \times \frac{|R_l|}{m} \tag{1.6}$$

The final privacy score based on Srivastava is given by:

$$PQ(j) = \sum_k \beta_k V(k,l) \tag{1.7}$$

Where the range of attributes can be between $[1,m]$. The final privacy score gained from the calculation indicates the potential privacy risks of the texts which have been shared and published by users in online social networks.

### 1.3.3.4  Privacy-Functionality Score

Domingo-Ferrer et al. [44] proposed a method for measuring the privacy risk of users in social networking sites. In a more specific manner, they tried to understand the benefits that users may obtain by sharing their information in online social networks, such as LinkedIn. For doing so, they computed the effectiveness that an individual may gain from sharing his/her information in social networks given by:

$$PRF(j) = \frac{\sum_{j'=1, j' \neq j}^{N} \sum_{i=1}^{n} \sum_{k=1}^{l} \beta_{ik} V(i,j',k) I(j,j',k)}{1 + PR_j} \tag{1.8}$$

$$PRF(j) = \frac{\sum_{j'=1, j' \neq j}^{N} \sum_{i=1}^{n} \sum_{k=1}^{l} \beta_{ik} V(i,j',k) I(j,j',k)}{1 + \sum_{i=1}^{n} \sum_{k=1}^{l} \beta_{ik} V(i,j,k)} \tag{1.9}$$

With the following conditions:

- $I(j,j',k) = 1$ If $j'$ and $j$ are $k$ links away from each other
- 0 otherwise

Where $j$ and $j'$ are the users in the social networks, $k$ indicates the number of links between users and $n$ indicates the number of attributes for a user. By applying this formula, users become able to decide what to share with other users at a specific time while maintaining the effectiveness of shared information in the desired social network.

### 1.3.3.5  SONET: Privacy Monitoring and Ranking

Nepali and Wang [45] introduced a monitoring and ranking model for privacy in the online social network. Six components formed their SONET model. These components comprise the model for the social network, browse and deduce data by data aggregation, privacy index (which is known as PIDX) and privacy invasion, privacy-preserving and security protection, and finally, monitoring and countermeasures that can be done regarding privacy breaches. The main component of the model is the privacy index where it computes the privacy exposure of individuals based on sensitivity and visibility. Based on Nepali and Wang, the privacy index (PIDX) is a factor of an entity's privacy vulnerability based on recognized properties. The final calculated model for the privacy index is between 0 and 100, given by:

$$PIDX = \left( \frac{W_{L_k}}{W_l} \right) \times 100 \tag{1.10}$$

Where $W_{L_k}$ is the summation of visibility of each attribute in its corresponding weight. Nepali and Wang define the privacy invasion as when $PIDX \geq T$, where $T$ indicates the threshold of security and privacy. If the computed index is higher than $T$, it means that the privacy of the entity is not preserved. Accordingly, the SONET model enables users to monitor their privacy index by the defined threshold for their privacy.

### 1.3.3.6    Privometer

Talukder et al. [46] developed the Privometer tool to compute the leakage of users' sensitive information. The computed score is shown by a numerical value. After providing the computed score as a numerical value, the tool is able to suggest preventative actions, which are known as self-sanitization actions. One of the main features of Privometer is considering all shared information by users in online social networks rather than just considering publicly available data. In contrast, a drawback of the tool is that it is only applicable to social network sites that permit applications to acquire and retrieve individuals' information.

### 1.3.3.7    PScore: Privacy Awareness Enhancement

Petkos et al. [40] proposed a framework to increase the understanding and knowledge of individuals who use online social networks, considering privacy and security. Three separate features have been considered for the framework: personal preference of users about the sensitivity of their information, the classified structure of information (attributes, values, and number of dimensions) to create a semantic organization, and inferred information out of the scope of social networks. These three features cover the prerequisites of the framework and clearly make a distinction of the framework compared with other scoring models.

Beside privacy scoring frameworks, some studies consider privacy settings in the online social network and proposed tools for better privacy configuration. Fang and LeFevre [47] proposed a "privacy wizard" using machine-learning techniques. Mazzia et al. [48] proposed a graphical tool that provides options for privacy, called as PViz. Although these scoring frameworks are able to measure and compute the privacy risk of users, the tools for privacy configuration can also help users and individuals to select the best settings for the privacy of information in their profile.

### B. Polytomous Approach

The other approach for calculation of privacy is polytomous-based. The difference of this approach with dichotomous one is in the level of shared information by individuals. As mentioned earlier, in the dichotomous approach, there are only two forms for shared information (the information is public or private), while in polytomous approach, we have more than two status for shared information by individuals. As this approach can obtain more different states for shared information, it can provide a more accurate estimation of calculated privacy for the online social network users. Liu and Terzi proposed a polytomous approach based on IRT. Becker et al. [39] discuss the significance of quantifying privacy in the online social network. They mentioned that quantifying privacy becomes even more critical in the case of protecting the huge volume of corresponding personal information, especially in large-scale online social networks. Finally, Aghasian et al. [38] proposed a privacy scoring model for multiple online social networks.

### 1.3.3.8   IRT-Based Privacy Risk Score

The first polytomous model has proposed by Liu and Terzi [41]. They expanded previous privacy score models by considering both the dichotomous and polytomous approaches. In their approach, they use the IRT-based method to compute a user's privacy. The derivation of Item Response Theory (IRT) in psychometrics returns to data analysis of gathered information from tests or questionnaire. In their model, a two-parameters (users and users' attributes) IRT-based model had been considered. In this case, their model can quantify users' concern about their privacy. Finally, their model can calculate and measure privacy score as mentioned in [42]. In their polytomous approach, they defined $R(k, l) = n$, indicates that the $k - th$ attribute for user $l$ has been disclosed to $n$ nodes further in a graph. Hence, the privacy score provides an indicator of whether the information of an individual is at risk or not. Considering Liu and Terzi's model for computing the privacy risk score for online social networks users, Sramka [49] extended the model to evaluate and assess the risk of privacy from the users' viewpoint. In their new privacy score, Sramka considered users' background knowledge that was publicly available. The main drawback of this method was that only attribute and identity revelation were measured and only a dichotomous approach was considered.

### 1.3.3.9   PrivAware - Quantifying Privacy Model

Becker et al. [39] presented a tool called PrivAware that is able to discover and report an accidental loss of information in online social networks. For creating the tool, they defined the problem as follows: friends' association with user $x$ have constituted as a set of data which form a record including type, value and weight. By defining a function which shows the value allocated to the user $x$'s friends, they were able to make the effective use of the outcomes with the support of friends with greater social value. As an example, the term type will contain values such as a university, zip code or age. Similarly, for each type of an attribute, there is a value which comes with the weight of that attribute. The weight here decides the importance of that attribute. It should be noted that the value of term weight has a value between zero and one that is set to the attributes based on the disambiguation process. In order to accomplish such task, they recruited 105 participants who shared their profile information for analysis. After proposing the inference algorithm, [39] they used three measures to assess the efficiency of the algorithm. The measures include Inferred attributes, verifiable inference, and correct inference.[5] They understood that the structured attributes have a tendency to be accurately inferred in a more time-consuming manner - such as country, age, high school graduation year, a state with an exception for zip code. Contrariwise, unstructured and semi-structured data tend to be further challenging to infer correctly.

---

[5]Inferred attributes indicate those attributes that are inferred by an algorithm. Verifiable inferences indicate that the inferred attributes by proposed algorithms are also available in the target user's profile.

### 1.3.3.10    Scoring Users' Privacy Disclosure Across Multiple Online Social Networks

Aghasian et al. [38] proposed a polytomous scoring framework to calculate a user's privacy disclosure score in multiple online social networks. In this regard, they considered two factors that impact a user's privacy: visibility and sensitivity. For measuring the visibility of information, three factors have been identified, namely accessibility, difficulty of data extraction and reliability. Accessibility refers to how available a user's piece of information is in online social networks—it indicates whether user information is publicly available, semi-public or completely private. Difficulty of data extraction relates to the amount of effort required to gain that information, while the reliability indicates if the obtained information is valid and reliable or not. After calculating visibility and gaining the value for sensitivity from prior studies, a fuzzy-based mathematical system has applied to measure the final privacy disclosure score of users.

Table 1.4 summarizes the various privacy scoring frameworks.

*Table 1.4    Comparison of historical studies*

| Focus | No. of Sources | Data Type | Approach | Reference |
|---|---|---|---|---|
| Attribute inference | 1 | Structured | Polytomous | [39] |
| Obtained utility by sharing information | 1 | Structured | Dichotomous | [44] |
| Sensitive information leakage of a profile | 1 | Structured | Dichotomous | [46] |
| Privacy risk from individual perspective | 1 | Structured | Dichotomous & Polytomous | [41] |
| Privacy risk of text messages | 1 | Unstructured | Dichotomous | [43] |
| Privacy exposure based on known parameters | - | Web data | Dichotomous | [45] |
| Privacy disclosure score of user's information | 4 | Structured | Polytomous | [38] |

So, accuracy and precision of the inferred attributes can be verified. Correct inferences indicate that the inferred attributes are correct matches of the attribute values in the target user's profile.

### 1.3.4   Privacy-preserving Approaches

Current research on privacy protection can be classified into three groups [50], each category concentrating on an aspect of privacy. Privacy by *policy* considers access models, privacy by *statistics* focuses on analytical and statistical technologies to generate tuned information revelation mechanisms, and privacy by *cryptography* aims to develop systems to guarantee the goal of CIA. While the information utility in privacy by the policy is high, there is no guarantee of privacy-preservation and the strength of privacy is low. In the second category, privacy by statistics, information utility and the strength of privacy is medium but its strength can be enhanced by particular patterns for accessing data. The third category, privacy by cryptography, has very high privacy strength and can guarantee the privacy in a theoretical manner which leads to low data and information utility. Beside privacy protection, data anonymisation has been broadly studied and commonly implemented for preserving data secrecy in non-collaborating data sharing and publishing scenarios [51]. On the other hand, data distribution with a great amount of individuals must consider several matters, involving efficiency, data integrity and secrecy of the data owner.

### 1.3.5   Privacy-preserving Models

Anonymising data relies on eliminating or altering the identifying variable(s) contained in the data, also known as personally identifiable information (PII).[6] Anonymising data keeps the referenced person's privacy as a priority while giving attention to a data analyzer needs (e.g., an analyst examining the data for identification of trends, patterns, etc.) [52]. Moreover, anonymisation is one of the common methods of providing sanitized data.[7] In this process, information which is identifiable is detached and other attributes are perturbed. Still, there are no assurances for stopping an attack from an intruder and attacker who has background knowledge of the data. Therefore, there is a need for providing other procedures and methods for data retrieval to preserve the privacy of the published information of user profiles in online social networks[32]. From the personalization perspective, diverse users may have dissimilar privacy preferences. As a case in point, some records are more important for some individuals, while other users may pay attention to other attributes. The focus of current methods in personalized protection is on "personalized access control" (e.g., attribute-based encryption [55]) or sensitivity personalization of a single dimension [56], while no one has explored sensitivity personalization in multidimensional data [57].

A variety of anonymisation algorithms with dissimilar anonymisation processes have been proposed by different authors [58, 56, 59, 60, 51, 61, 62, 63]. Models like *k*-anonymity, *l*-diversity and *t*-closeness are the most approved and accepted methods that deliver appropriate outcomes in anonymisation. *K*-anonymity [63] and *l*-diversity [61] are the main accepted models on privacy to quantify the degree of pri-vacy, for sensitive information revelation against record linkage attack and attribute linkage attacks, respectively. Supplementary secrecy models such as *t*-closeness [64] and *m*-invariance [65] are also presented for numerous attack in privacy scenarios. Numerous anonymising processes are applied to maximize the advantage of anonymise data-sets, as well as suppression [66], generalization [51, 60], anatomisation [67], slicing [68], disassociation [69].

---

[6]Typically, an identifying variable is one that defines an attribute of an individual that is visible and evident, which is recorded (such as social security number, employee ID, patient ID, etc.), or other people can identify.

[7]Data sanitization is the procedure of veiling sensitive data and create datasets by overwriting it with accurate but incorrect information of an identical type [53, 54].

### 1.3.5.1   *k*-anonymity

*k*-anonymity is the most common method in privacy-preservation against record linkage attack. If the information for each individual stays undistinguished for the other *k*-1 individuals in a data-set, then the *k*-anonymity is fulfilled. This can be done by two different methods:

- Suppression: removing the value of an attribute from the perturbed data.
- Generalization: substitution of an attribute with a less detailed but semantically reliable value.

  Moreover, different classifications of attributes in *k*-anonymity should be considered[63].

- Key attribute: a user can be identified directly by this attribute.
- Quasi-Identifier (*QI*) : provide capability to recognize a user by a set of parameters and attributes.
- Sensitive attribute.

This method can guarantee that users are safe from linking attacks while they may not be secured and safeguarded against attribute revelation. For example, in order to perform a generalization of a value with suppression for ethnicity, three different levels may exist. At the first level, three different ethnics may appear including Asian, European and South American. At the second level of generalization, only person class is exists for all three ethnic groups and at the last level, every record is anonymised and nothing is available to other users or an adversary. Table 1.5 shows an example of anonymised data using *k*-anonymity method.

*Table 1.5   An example of k-anonymised data*

| Nationality | Zip code | Disease |
|:---:|:---:|:---|
| * | 878XX | Acne |
| * | 878XX | Acne |
| * | 878XX | Flu |
| * | 878XX | Flu |

Two techniques are applied to achieve *k*-anonymity and hiding information, the suppression method (applied to the nationality attribute), which does not publish any value, and the generalization method (applied to the zip code attribute), which uses consistent but less specific values to perform anonymisation.

Till now, we have discussed how *k*-anonymity methods work for preventing the privacy. Like other methods, *k*-anonymity has some drawbacks and potential attacks may still occur on the anonymised data, including [61]:

- Homogeneity attack, when there is insufficient diversity for the sensitive parameters in a quotient space (equivalence class)[8] of the dataset.
- Background knowledge attack, when the adversary has contextual information and facts.

A privacy breach on the anonymised data can happen if one of the mentioned attacks occur. In order to protect the information against these attacks, other methods have been proposed which are discussed in the next section.

### 1.3.5.2    *l*-diversity

*L*-diversity is a method that can lessen the risks in *k*-anonymity regarding the revelation of sensitive information [61]. It guarantees that the values of sensitive parameters are dissimilar in each equivalence class. While *l*-diversity enhances the privacy preservation compared with *k*-anonymity and helps to mitigate the risks that may occur when using *k*-anonymity, it is still likely that an adversary infers sensitive information. This can happen if the distribution of a sensitive record in a cluster is very different from the distribution of identical attributes in that class.

Li et al. [70] showed two possible attacks on *l*-diversity including skewness attack and similarity attack and indicated that the *l*-diversity method cannot safeguard against these sorts of attacks. Table 1.6 shows that sensitive information needs to be diverse in each *QI* equivalence class. As each sensitive attribute (disease) is diverse within each quotient space, no individual may be re-identified.

*Table 1.6    An example of l-diversity table*

| Nationality | Zip code | Disease |
|-------------|----------|---------|
| Asian | 878XX | Acne |
| Asian | 878XX | Flu |
| Asian | 878XX | Acne |
| Asian | 878XX | Shingles |
| Asian | 878XX | Flu |
| Asian | 878XX | Flu |
| American | 878XX | Flu |
| American | 878XX | Acne |
| American | 878XX | Flu |
| American | 878XX | Flu |
| American | 878XX | Acne |
| American | 878XX | Shingles |

Machanavajjhala et al. [61] describes a variety of *l*-diversity techniques, including:

1. Distinct *l*-diversity: bounding the occurrence of the most frequent value by $1/l$ in an equivalence class.

---

[8] A equivalence class is a set of clusters.

2.  Entropy $l$-diversity: $log(l)$ is the least acceptable entropy of the sensitive information distribution in each equivalence class.
3.  Recursive $(c, l)$-diversity: the most common value does not appear regularly.

### 1.3.5.3    $T$-closeness

In order to solve the problems of previous methods for preventing the privacy, Li et al. [64, 70] proposed an intuitive privacy preserving model named $t$-closeness. They indicate that the sensitive information distribution within each $QI$ compared with its distribution in the original dataset should be close. They also proposed an $(n, t)$-closeness privacy method which is more flexible [64]. In this case, the method bounds the number of released sensitive attributes that an adversary or an observer can gain from the table.

### 1.3.5.4    Differential privacy

The concept of differential privacy was initially proposed by Dwork [71] which safeguards private distinguishable data at the severest probable level. It addresses the situation when a reliable data custodian desires to publish some statistics over its data, devoid of disclosing information about a specific value itself. This is done by adding noise to a small sample of user's usage pattern. Dwork defined differential privacy as "A randomized algorithm $M$ is $\varepsilon$-differentially private if for all pairs of adjacent databases $x, y$, and for all sets $S \subset Range(M(x)) \cup Range(M(y))$":

$$Pr[M(x) \in S] \leq e^{\varepsilon} \cdot Pr[M(y) \in S]$$

where the probabilities are over algorithm $M'$s coins, $e$ stands for exp and '·' indicates that the transformation is stable in at most $\varepsilon$-times of the hamming distance between two data sets.

Table 1.7 summarizes the most common methods of preserving privacy for data publishing.

## 1.4    Privacy Preservation in Recommender Systems

Different methods have been proposed for preserving the privacy of users in recommender systems. Badsha et al. [79] presented a pragmatic privacy-preserving content-based filtering recommender system which works based on homomorphic encryption. To achieve this, they calculate item to item similarity of one user and then generate secure recommendations (provide recommendations without revealing sensitive information of users). Nikolaenko et al. [80] proposed a new method to leverage sparsity of data to achieve security in recommender systems, which works based on matrix factorization. Shokri et al. [81] developed a distributed aggregation mechanism for individuals to obscure the connection between item and user in data sent to a server that is not trusted in a collaborative filtering based recommender system. They guarantee that the privacy of users will be kept while the least information loss occurs for individuals. Machanavajjhala et al. [82] proposed a differential based privacy-preserving method for graph-based social networks to increase the trade-off between data utility and privacy. In their model, they have considered all edges of the graph as sensitive and proposed an algorithm that was able to recommend a single node for a few target nodes. Hoffmann et al. [83] proposed a privacy scheme for collaborative filtering recommender systems by factor analysis. They also make use of a peer-to-peer protocol to meet the privacy of individuals' information in their model. [84, 85, 86, 87, 88] developed and proposed privacy-preserving algorithms and models for recommender systems. In their models, they tried to increase the efficiency of algorithms or increase the effectiveness of accuracy versus privacy in such systems. Table 1.8 presents an overview of common recommender-based systems privacy preservation models.

*Table 1.7   An overview of common privacy preserving techniques*

| Presented Model | Applied Technique | Author(s) | Year |
|---|---|---|---|
| $K$-minimal generalization | Domain generalization hierarchies of the *QI* | Samarati [72] | 2001 |
| Simple $k$-anonymity model | Mapping information to no, $k$ or incorrect entities | Sweeney [63] | 2002 |
| Bottom-up generalization | Masking and hiding information instead of learn patterns | Wang et al. [73] | 2004 |
| Top-Down Specialization | Specifying the level of information in a top-down way till the least privacy condition is disrupted | Fung et al. [74] | 2005 |
| Enhanced $k$-anonymity model | Protecting both identifications and relationships to sensitive information in data | Wong et al. [75] | 2006 |
| $K$-anonymity in classification issues | Suppression and Progressive Disclosure Algorithm | Fung et al. [58] | 2007 |
| Differential privacy | Adding properly random noise in data | Dwork [71] | 2008 |
| $P$-sensitive $k$-anonymity privacy model | Modifying the initial *QI* attributes values | Sun et al. [76] | 2008 |
| Enhanced $(L, \alpha)$-diversity | Controlling the weight of sensitive information in a in a given quasi identifier cluster | Sun et al. [77] | 2011 |
| Slicing | Slicing the dataset in a vertical and horizontal manner | Li et al [68] | 2012 |
| Concentrated differential privacy | Adding properly random noise in data | Dwork & Rothblum [78] | 2016 |

*Table 1.8    An overview of common recommender-based systems privacy preservation models*

| Applied Technique | Author(s) | Year |
|---|---|---|
| Probabilistic factor (correlation and regression) analysis | Hoffmann et al. [83] | 2005 |
| Distributed aggregation method by modelling a bipartite graph | Shokri et al. [81] | 2009 |
| Link-analysis of graph based on the differential privacy and Laplace and exponential smoothing algorithms | Machanava-jjhala et al. [82] | 2011 |
| Matrix factorisation (collaborative filtering) | Nikolaenko et al. [80] | 2013 |
| Data perturbation by micro-aggregation | Casino et al. [86] | 2015 |
| Homomorphic encryption based on ElGamal crypto-system | Badsha et al. [79] | 2016 |

## 1.5    Conclusion and Future Directions

The growing use of recommender systems in online social networks presents a privacy risk for the many users of these networks. The risks can be considered from two different perspectives: measuring a user's risk of unintended information disclosure and techniques to preserve users' privacy when sharing large datasets. The general objective of the first perspective is to understand both dichotomous and polytomous approaches and the differences between them for measuring privacy. From the privacy-preservation perspective, the challenges in protecting the privacy of online social networks users and recommender systems are to propose real-time methods which can support a high volume of data with data types. Then, we discussed the privacy-preservation models (anonymisation techniques and encryption-based methods) for the users in these two systems which all provide information sanitization and data obfuscation to assure data anonymity of individuals.

As users' participation in different types of online social media and recommender systems are increasing rapidly, privacy-preservation of individuals is be-coming more challenging. Hence, there is a need to propose new privacy preservation models in the near future that can deal with different data types, and for privacy-preservation systems to take into account dependencies between data. Another consideration is resource consumption. As providing privacy requires substantial computation, factors that impact on computation time should be studied to identify whether new mechanisms are required that are less computationally intensive. Modeling attacks is another significant issue that should be taken into account. While different types of attacks occur on anonymised datasets, modeling attacks on datasets with more complicated features should be a priority so that vulnerabilities can be uncovered before they are exploited. Finally, there is a need to propose novel methods to help organizations to preserve the privacy of users while these organizations are storing, analyzing and mining individuals' data within their organizations.

# Bibliography

[1] Gross R, Acquisti A. Information revelation and privacy in online social networks. In: Proceedings of the 2005 ACM workshop on Privacy in the electronic society. ACM;. p. 71–80.

[2] Ahmadizadeh E, Aghasian E, Taheri HP, et al. An Automated Model to Detect Fake Profiles and botnets in Online Social Networks Using Steganography Technique. IOSR Journal of Computer Engineering (IOSR-JCE). 2015;17:65–71.

[3] Jeckmans AJ, Beye M, Erkin Z, et al. Privacy in recommender systems. In: Social media retrieval. Springer; 2013. p. 263–281.

[4] Backstrom L, Dwork C, Kleinberg J. Wherefore art thou r3579x?: anonymized social networks, hidden patterns, and structural steganography. In: Proceedings of the 16th international conference on World Wide Web. ACM;. p. 181–190.

[5] Ricci F, Rokach L, Shapira B. Introduction to recommender systems handbook. In: Recommender systems handbook. Springer; 2011. p. 1–35.

[6] Resnick P, Varian HR. Recommender Systems. Commun ACM. 1997 Mar;40(3):56–58. Available from: http://doi.acm.org/10.1145/245108.245121.

[7] Billsus D, Pazzani MJ. User modeling for adaptive news access. User modeling and user-adapted interaction. 2000;10(2-3):147–180.

[8] Burke R. Hybrid recommender systems: Survey and experiments. User modeling and user-adapted interaction. 2002;12(4):331–370.

[9] Rich E. User modeling via stereotypes. Cognitive science. 1979;3(4):329–354.

[10] Trewin S. Knowledge-based recommender systems. Encyclopedia of library and information science. 2000;69(Supplement 32):180.

[11] Aldhafferi N, Watson C, Sajeev AS. Personal information privacy settings of online social networks and their suitability for mobile internet devices [Journal Article]. arXiv preprint arXiv:13052770. 2013;.

[12]   Beye M, Jeckmans AJP, Erkin Z, et al. In: Privacy in online social networks. Springer; 2012. p. 87–113.

[13]   Bnnig C, Cap CH. Ad hoc privacy management in ubiquitous computing environments. In: Advances in Human-oriented and Personalized Mechanisms, Technologies, and Services, 2009. CENTRIC'09. Second International Conference on. IEEE;. p. 85–90.

[14]   Kang J. Information privacy in cyberspace transactions [Journal Article]. Stanford Law Review. 1998;p. 1193–1294.

[15]   Ni Q, Bertino E, Lobo J, et al. Privacy-aware role-based access control [Journal Article]. ACM Transactions on Information and System Security (TISSEC). 2010;13(3):24.

[16]   Solove DJ. A taxonomy of privacy [Journal Article]. University of Pennsylvania law review. 2006;p. 477–564.

[17]   Stone EF, Gueutal HG, Gardner DG, et al. A field experiment comparing information-privacy values, beliefs, and attitudes across several types of organizations [Journal Article]. Journal of applied psychology. 1983;68(3):459.

[18]   Taheri S, Hartung S, Hogrefe D. Achieving receiver location privacy in mobile ad hoc networks. In: Social Computing (SocialCom), 2010 IEEE Second International Conference on. IEEE;. p. 800–807.

[19]   Smith HJ, Dinev T, Xu H. Information privacy research: an interdisciplinary review [Journal Article]. MIS quarterly. 2011;35(4):989–1016.

[20]   Whitman ME, Mattord HJ. Principles of information security. Cengage Learning; 2011.

[21]   Cherdantseva Y, Hilton J. A reference model of information assurance & security. In: Availability, reliability and security (ares), 2013 eighth international conference on. IEEE;. p. 546–555.

[22]   Kaplan AM, Haenlein M. Users of the world, unite! The challenges and opportunities of Social Media [Journal Article]. Business horizons. 2010;53(1):59–68.

[23]   Boyd D, Ellison N. social network sites: definition, history, scholarship: Department of Telecommunication [Journal Article]. Information Studies, and Media, Michigan State University. 2009;.

[24]   Stan J, Muhlenbach F, Largeron C. Recommender systems using social network analysis: Challenges and future trends. In: Encyclopedia of Social Network Analysis and Mining. Springer; 2014. p. 1522–1532.

[25]   Zeadally S, Badra M. Privacy in a Digital, Networked World [Journal Article]. 2015;.

[26]   Raynes-Goldie KS. Privacy in the age of Facebook: Discourse, architecture, consequences. Curtin University.; 2012.

[27]   Palen L, Dourish P. Unpacking privacy for a networked world. In: Proceedings of the SIGCHI conference on Human factors in computing systems. ACM; 2003. p. 129–136.

[28]   Leenes R. Context is everything sociality and privacy in online social network sites. In: IFIP PrimeLife International Summer School on Privacy and Identity Management for Life. Springer; 2009. p. 48–65.

[29]    Rosenblum D. What anyone can know: The privacy risks of social networking sites [Journal Article]. IEEE Security and Privacy. 2007;5(3):40–49.

[30]    Bonneau J. Attack of the zombie photos [Journal Article]. Light Blue Touchpaper http://www lightbluetouchpaper org/2009/05/20/attackof-the-zombie-photos. 2009;.

[31]    Walters C. Facebook's new terms of service:" We can do anything we want with your content. Forever." [Journal Article]. The Consumerist. 2009;15.

[32]    Zheleva E, Terzi E, Getoor L. Privacy in social networks [Journal Article]. Synthesis Lectures on Data Mining and Knowledge Discovery. 2012;3(1):1–85.

[33]    Lenhart A. Teens, Online Stranger Contact & Cyberbullying: What the Research is Telling Us. Pew Internet & American Life Project; 2008.

[34]    Madden M, Lenhart A, Cortesi S, et al. Teens, social media, and privacy [Journal Article]. Pew Research Center. 2013;21.

[35]    Kowalski RM. I was only kidding Victims and perpetrators perceptions of teasing [Journal Article]. Personality and Social Psychology Bulletin. 2000;26(2):231–241.

[36]    Madden M, Smith A. Reputation management and social media [Journal Article]. 2010;.

[37]    Renner C. Privacy in Online Social Networks [Thesis]; 2010.

[38]    Aghasian E, Garg S, Gao L, et al. Scoring Users Privacy Disclosure Across Multiple Online Social Networks. IEEE Access. 2017;.

[39]    Becker JL, Chen H. Measuring privacy risk in online social networks [Thesis]; 2009.

[40]    Petkos G, Papadopoulos S, Kompatsiaris Y. PScore: A Framework for Enhancing Privacy Awareness in Online Social Networks. In: Availability, Reliability and Security (ARES), 2015 10th International Conference on. IEEE;. p. 592–600.

[41]    Liu K, Terzi E. A framework for computing the privacy scores of users in online social networks [Journal Article]. ACM Transactions on Knowledge Discovery from Data (TKDD). 2010;5(1):6.

[42]    Maximilien EM, Grandison T, Liu K, et al. Enabling privacy as a fundamental construct for social networks. In: Computational Science and Engineering, 2009. CSE'09. International Conference on. vol. 4. IEEE;. p. 1015–1020.

[43]    Srivastava A, Geethakumari G. Measuring privacy leaks in online social networks. In: Advances in Computing, Communications and Informatics (ICACCI), 2013 International Conference on. IEEE;. p. 2095–2100.

[44]    Domingo-Ferrer J. Rational privacy disclosure in social networks. In: International Conference on Modeling Decisions for Artificial Intelligence. Springer; 2010. p. 255–265.

[45]    Nepali RK, Wang Y. Sonet: A social network model for privacy monitoring and ranking. In: 2013 IEEE 33rd International Conference on Distributed Computing Systems Workshops. IEEE;. p. 162–166.

[46]  Talukder N, Ouzzani M, Elmagarmid AK, et al. Privometer: Privacy protection in social networks. In: Data Engineering Workshops (ICDEW), 2010 IEEE 26th International Conference on. IEEE;. p. 266–269.

[47]  Fang L, LeFevre K. Privacy wizards for social networking sites. In: Proceedings of the 19th international conference on World wide web. ACM; 2010. p. 351–360.

[48]  Mazzia A, LeFevre K, Adar E. The PViz comprehension tool for social network privacy settings. In: Proceedings of the Eighth Symposium on Usable Privacy and Security. ACM; 2012. p. 13.

[49]  Sramka M. In: Evaluating Privacy Risks in Social Networks from the Users Perspective. Springer; 2015. p. 251–267.

[50]  Zhang L, Li XY, Lei J, et al. Mechanism design for finding experts using locally constructed social referral web [Journal Article]. Parallel and Distributed Systems, IEEE Transactions on. 2015;26(8):2316–2326.

[51]  Fung B, Wang K, Chen R, et al. Privacy-preserving data publishing: A survey of recent developments [Journal Article]. ACM Computing Surveys (CSUR). 2010;42(4):14.

[52]  Abuelsaad TE, Hoyos C. Data Perturbation and Anonymization Using One Way Hash [Generic]. Google Patents; 2011.

[53]  Tambe P, Vora D. Data sanitization for privacy preservation on Social Network. In: Automatic Control and Dynamic Optimization Techniques (ICAC-DOT), International Conference on. IEEE; 2016. p. 972–976.

[54]  Edgar D. Data sanitization techniques. A Net. 2000;p. 2003–2004.

[55]  Li M, Yu S, Zheng Y, et al. Scalable and secure sharing of personal health records in cloud computing using attribute-based encryption [Journal Article]. Parallel and Distributed Systems, IEEE Transactions on. 2013;24(1):131–143.

[56]  Xiao X, Tao Y. Personalized privacy preservation. In: Proceedings of the 2006 ACM SIGMOD international conference on Management of data. ACM;. p. 229–240.

[57]  Wang W, Chen L, Zhang Q. Outsourcing high-dimensional healthcare data to cloud with personalized privacy preservation [Journal Article]. Computer Networks. 2015;88:136–148.

[58]  Fung B, Wang K, Yu PS. Anonymizing classification data for privacy preservation [Journal Article]. Knowledge and Data Engineering, IEEE Transactions on. 2007;19(5):711–725.

[59]  LeFevre K, DeWitt DJ, Ramakrishnan R. Incognito: Efficient full-domain k-anonymity. In: Proceedings of the 2005 ACM SIGMOD international conference on Management of data. ACM;. p. 49–60.

[60]  LeFevre K, DeWitt DJ, Ramakrishnan R. Mondrian multidimensional k-anonymity. In: Data Engineering, 2006. ICDE'06. Proceedings of the 22nd International Conference on. IEEE;. p. 25–25.

[61]  Machanavajjhala A, Kifer D, Gehrke J, et al. l-diversity: Privacy beyond k-anonymity [Journal Article]. ACM Transactions on Knowledge Discovery from Data (TKDD). 2007;1(1):3.

[62]    Samarati P, Sweeney L.  Protecting privacy when disclosing information: k-anonymity and its enforcement through generalization and suppression. Technical report, SRI International; 1998.

[63]    Sweeney L. k-anonymity: A model for protecting privacy [Journal Article]. International Journal of Uncertainty, Fuzziness and Knowledge-Based Systems. 2002;10(05):557–570.

[64]    Li N, Li T, Venkatasubramanian S.  Closeness: A new privacy measure for data publishing [Journal Article]. IEEE Transactions on Knowledge and Data Engineering. 2010;22(7):943–956.

[65]    Xiao X, Tao Y.  M-invariance: towards privacy preserving re-publication of dynamic datasets. In: Proceedings of the 2007 ACM SIGMOD international conference on Management of data. ACM;. p. 689–700.

[66]    Wang K, Fung BCM, Philip SY.  Handicapping attacker's confidence: an alternative to k-anonymization [Journal Article]. Knowledge and Information Systems. 2007;11(3):345–368.

[67]    Xiao X, Tao Y.  Anatomy: Simple and effective privacy preservation. In: Proceedings of the 32nd international conference on Very large data bases. VLDB Endowment; 2006. p. 139–150.

[68]    Li T, Li N, Zhang J, et al.  Slicing: A new approach for privacy preserving data publishing [Journal Article]. IEEE transactions on knowledge and data engineering. 2012;24(3):561–574.

[69]    Terrovitis M, Mamoulis N, Liagouris J, et al.  Privacy preservation by disassociation [Journal Article].  Proceedings of the VLDB Endowment. 2012;5(10):944–955.

[70]    Li N, Li T, Venkatasubramanian S. t-closeness: Privacy beyond k-anonymity and l-diversity. In: Data Engineering, 2007. ICDE 2007. IEEE 23rd International Conference on. IEEE; 2007. p. 106–115.

[71]    Dwork C. An ad omnia approach to defining and achieving private data analysis. In: Privacy, Security, and Trust in KDD. Springer; 2008. p. 1–13.

[72]    Samarati P.  Protecting respondents identities in microdata release.  IEEE transactions on Knowledge and Data Engineering. 2001;13(6):1010–1027.

[73]    Wang K, Yu PS, Chakraborty S.  Bottom-up generalization: A data mining solution to privacy protection.  In: Data Mining, 2004. ICDM'04. Fourth IEEE International Conference on. IEEE; 2004. p. 249–256.

[74]    Fung BCM, Wang K, Yu PS.  Top-down specialization for information and privacy preservation. In: 21st International Conference on Data Engineering (ICDE'05). IEEE;. p. 205–216.

[75]    Wong RCW, Li J, Fu AWC, et al.  ($\alpha$, k)-anonymity: an enhanced k-anonymity model for privacy preserving data publishing.  In: Proceedings of the 12th ACM SIGKDD international conference on Knowledge discovery and data mining. ACM; 2006. p. 754–759.

[76]    Sun X, Wang H, Li J, et al.  Enhanced p-sensitive k-anonymity models for privacy preserving data publishing. Trans Data Privacy. 2008;1(2):53–66.

[77]  Sun X, Li M, Wang H.  A family of enhanced (L, $\alpha$)-diversity models for privacy preserving data publishing.  Future Generation Computer Systems. 2011;27(3):348–356.

[78]  Dwork C, Rothblum GN.  Concentrated differential privacy.  arXiv preprint arXiv:160301887. 2016;.

[79]  Badsha S, Yi X, Khalil I.  A practical privacy-preserving recommender system.  Data Science and Engineering. 2016;1(3):161–177.

[80]  Nikolaenko V, Ioannidis S, Weinsberg U, et al.  Privacy-preserving matrix factorization.  In: Proceedings of the 2013 ACM SIGSAC conference on Computer & communications security. ACM; 2013. p. 801–812.

[81]  Shokri R, Pedarsani P, Theodorakopoulos G, et al.  Preserving privacy in collaborative filtering through distributed aggregation of offline profiles. In: Proceedings of the third ACM conference on Recommender systems. ACM; 2009. p. 157–164.

[82]  Machanavajjhala A, Korolova A, Sarma AD.  Personalized social recommendations:  accurate or private.  Proceedings of the VLDB Endowment. 2011;4(7):440–450.

[83]  Hofmann T, Hartmann D. Collaborative filtering with privacy via factor analysis.  In: Proceedings of the 2005 ACM symposium on applied computing; 2005. p. 791–795.

[84]  Li D, Chen C, Lv Q, et al.  An algorithm for efficient privacy-preserving item-based collaborative filtering.  Future Generation Computer Systems. 2016;55:311–320.

[85]  Boutet A, Frey D, Guerraoui R, et al. Privacy-preserving distributed collaborative filtering. Computing. 2016;98(8):827–846.

[86]  Casino F, Domingo-Ferrer J, Patsakis C, et al.  A k-anonymous approach to privacy preserving collaborative filtering. Journal of Computer and System Sciences. 2015;81(6):1000–1011.

[87]  Jorgensen Z, Yu T. A Privacy-Preserving Framework for Personalized, Social Recommendations. In: EDBT; 2014. p. 571–582.

[88]  Shang S, Hui Y, Hui P, et al. Beyond personalization and anonymity: Towards a group-based recommender system.  In: Proceedings of the 29th Annual ACM Symposium on Applied Computing. ACM; 2014. p. 266–273.